\documentclass[prl,amssymb,twocolumn,showpacs]{revtex4}
\usepackage{graphicx}% Include figure files

\begin{document}

\title{Spin-orbit gap of graphene: First-principles calculations}
\author{Yugui Yao$^{1}$, Fei Ye$^{2}$, Xiao-Liang Qi$^{2}$, Shou-Cheng Zhang$^{3}$ and Zhong Fang$^{1,4}$}

\affiliation{$^1$ Beijing National Laboratory for Condensed Matter
Physics, Institute of Physics, Chinese Academy of Sciences,
Beijing, 100080, China}

\affiliation{$^2$ Center for Advanced Study, Tsinghua University,
Beijing, 100084, China}

\affiliation{$^3$ Department of Physics, McCullough Building,
Stanford University, Stanford, CA 94305-4045}

\affiliation{$^4$ International Center for Quantum Structure,
Chinese Academy of Sciences, Beijing, 100080, China}

\begin{abstract}
Even though graphene is a low energy system consisting of the two
dimensional honeycomb lattice of carbon atoms, its quasi-particle
excitations are fully described by the $2+1$ dimensional
relativistic Dirac equation. In this paper we show that while the
spin-orbit interaction in graphene is of the order of $4 meV$, it
opens up a gap of the order of $10^{-3} meV$ at the Dirac points.
We present the first principle calculation of the spin-orbit gap,
and explain the behavior in terms of a simple tight-binding model.
Our result also shows that the recently predicted quantum spin
Hall effect in graphene can only occur at unrealistically low
temperature.
\end{abstract}
\pacs{73.43.-f, 71.70.Ej, 73.21.-b} \maketitle

Recently, electronic properties of graphene, a single-layer
graphite sheet, attracted great interests both theoretically and
experimentally. The key difference of graphene compared with most
other two-dimensional materials is the linear energy spectrum
around two nodal points in Brillouin zone, which makes the low
energy dynamics of electrons in this system equivalent to that of
relativistic fermions, as described by the massless Dirac equation
\cite{semenoff1984}. The two sublattices in graphene honeycomb
lattice play the role of pseudo-spin degrees of freedom. In Ref.
[\onlinecite{novoselov2005,zhang2005}], the quantum Hall effect in
graphene is observed, which shows non-conventional quantization
rule $\sigma_H=\frac{2e^2}h(2n+1),n\in \mathbb{Z}$. Such an
"abnormal" quantum Hall effect agrees with the theoretical
calculation based on massless Dirac equation under external
magnetic field \cite{gusynin2005a,gusynin2005b,peres2006}, and can
be considered as a consequence of the chiral anomaly in two
dimensional massless fermions. Moreover, recent experiment on low
field magnetoresistance \cite{morozov2006} shows that graphene
remains metallic under temperature as low as $T=4K$, which
confirms that any possible gap opened at the Dirac cones cannot be
larger than $k_BT\sim 0.34 {\rm meV}$.

Nevertheless, it has been proposed that a small gap can open on
the two Dirac points of graphene due to spin-orbital coupling
(SOC) \cite{kane2005a}, which at the same time makes the system a
spin-Hall insulator \cite{murakami2004B} with quantized spin Hall
conductance. Physically, this proposal is a spinful version of
Haldane's model for quantum Hall effect without magnetic field
\cite{haldane1988}, in which a spin dependent
next-nearest-neighbor hopping term is introduced to induce
opposite mass terms for the two Dirac cones. Ref.
[\onlinecite{kane2005a}] estimates the spin-orbit gap in graphene
to be $2.4K$. In this paper we provide systematic calculations of
the spin-orbital gap in graphene by both first-principle
calculation and tight-binding model, which shows that the actual
gap is much smaller compared to the crude estimate given in Ref.
[\onlinecite{kane2005a}], it explains the (nearly) gaplessness
observed in experiments and also defines a much more narrow
temperature range for the quantum spin Hall effect to be observed.

The $sp^{2}$ hybridization of the $2s$ orbital and two $2p$
orbitals of carbon atom creates the $\sigma$ bonds to form the
honeycomb lattice of graphene which is bipartite with two carbon
atoms in one unit cell. The $\pi$ band consisting of the remaining
$2p$ orbitals controls the low energy physics of graphene and
makes it a semi-metal. One can describe the $\pi$ and $\sigma$
electrons by two tight binding (TB) Hamiltonians separately, which
in momentum space is a $2\times 2$ matrix
$\mathbf{H}_{\pi}(\vec{k})$ for $\pi$ band, and a $6\times 6$
matrix $\mathbf{H}_{\sigma} (\vec{k})$ for $\sigma$ band
\cite{saito1}. If the spin degeneracy of electrons is taken into
account, the dimension of these two matrices are doubled. The
diagonal entities of the matrices are the on-site energies of
different orbitals and the off-diagonal entities are the possible
hopping between different sublattices.

The SOC is a relativistic effect described by the Hamiltonian with
the form ${\hbar}\vec{\sigma}\cdot (\vec{\nabla} V \times \vec{p})
/(4m^{2}c^{2})\sim \vec{L}\cdot\vec{\sigma}$. $\vec{\sigma}$ is the
Pauli matrix. For a single carbon atom, there is no SOC between $2s$
and $2p$ orbitals due to their different azimuthal quantum number,
and the SOC only exists among the $2p$ orbitals. Its magnitude
$\xi_{0}$ can be estimated  of order $4meV$ by directly computing
the overlap integral of SOC between $2p_{z}$ and $2p_{x}$ orbitals.
Note that the SOC changes the magnetic quantum number accompanied
with the spin-flip of electrons, hence no SOC exists between the
same atomic orbital.

For the graphene, only the SOC in the normal direction with the form
$L_{z}\sigma_{z}$ has nonzero contribution due to the reflection
symmetry with respect to the lattice plane. Even this term vanishes
for the $\pi$ orbitals between nearest neighbors(NNs), since there
is an additional vertical reflection plane along the nearest
neighbor bond. Under the mirror reflection to this plane, the
$2p_{z}$ wavefunctions of the adjacent atoms are unchanged, however
the angular momentum $L_{z}$ changes its sign, hence the matrix
element of $L_{z}\sigma_{z}$ between NNs vanishes. This is different
from the carbon nanotube\cite{ando2000} where the curvature effect
could provide a SOC between $2p$ orbitals of NNs although it is
still vanishing for large tube radius. Thus, to realize the SOC
effect of $\pi$ band within the NNs approximation we need the aid of
$\sigma$-band. This process turns out to be a second order one,
which is three orders of magnitude smaller than $\xi_{0}$. On the
other hand, the SOC can act directly within the $\sigma$-band, it
will open a gap at some degenerate points with the same order as
$\xi_{0}\sim 4 meV$.

The SOC mixes the $\pi$ and $\sigma$ bands and the total
Hamiltonian reads
\begin{eqnarray}
  H=\left(
  \begin{array}{cc}
    \mathbf{H}_{\pi} & \mathbf{T}\\
    \mathbf{T}^{\dagger} & \mathbf{H}_{\sigma}
  \end{array}
  \right)\label{originalH}
\end{eqnarray}
Here, $\mathbf{H}_{\pi}$ and $\mathbf{H}_{\sigma}$ should be
enlarged to be $4\times 4$ and $12\times 12$ matrices by the spin
indices, respectively. The SOC term $\mathbf{T}$ bridging the
$\pi$ and $\sigma$ bands is a $4\times 12$ matrix of order
$\xi_{0}$, and its explicit form is not important at present, and
will be given later. The wavevector $\vec{k}$ is omitted for
simplicity hereinafter because it is always a good quantum number.

Since we are concerned with the low energy physics, an effective
$\pi$-band model with SOC derived from the original Hamiltonian
Eq.~(\ref{originalH}) is more advantageous. For this purpose, one
can perform a canonical transformation
\begin{eqnarray}
  &&H\rightarrow H_{S}=e^{-S}He^{S}\nonumber\\
  &&S=\left(
  \begin{array}{cc}
    0 & \mathbf{M}\\
    -\mathbf{M}^{\dagger} & 0
  \end{array}
  \right)
\end{eqnarray}
where $\mathbf{M}$ should satisfy
\begin{eqnarray}
  \mathbf{M}\mathbf{H}_{\sigma}-\mathbf{H}_{\pi}\mathbf{M}
  =\mathbf{T}
  \label{M}
\end{eqnarray}
so that $H_{S}$ is block diagonal up to order $\xi_{0}^{2}$.
Clearly $\mathbf{M}$ is also a $4\times 12$ matrix. The
effective Hamiltonian $H_{eff}$ is then extracted from the
diagonal part of $H_{S}$ as
\begin{eqnarray}
  H_{eff}\approx \mathbf{H}_{\pi}-\frac{1}{2}(\mathbf{T}\mathbf{M}^{\dagger}
  +\mathbf{M}\mathbf{T}^{\dagger})
\end{eqnarray}
The second term is just the effective SOC for the $\pi$ band
electrons.

The matrix $\mathbf{M}$ can be calculated iteratively through
Eq.~(\ref{M})
\begin{eqnarray}
  \mathbf{M}&=& \mathbf{T} \mathbf{H} ^{-1} _{\sigma}
  +\mathbf{H}_{\pi}\mathbf{T}\mathbf{H}^{-2}_{\sigma}+\cdots.
\end{eqnarray}
Around the Dirac points, the spectrum of $\mathbf{H}_{\pi}$ is
close to zero measured from the on-site potential of $2p$ orbital,
while that of $\mathbf{H} _{\sigma}$ is of order several $eV$,
hence we can take $\mathbf{M}\approx\mathbf{T}
\mathbf{H}^{-1}_{\sigma}$ approximately. The effective SOC of
$\pi$-band then reads
\begin{eqnarray}
  -\mathbf{T}\mathbf{H}_{\sigma}^{-1} \mathbf{T}^{\dagger}
  \label{soeffective}
\end{eqnarray}
whose magnitude is roughly estimated as $\xi_{1}\sim
|\xi_{0}|^{2}/\Delta$ with $\Delta$ being of the order of the
energy difference at the Dirac points between $\pi$ and $\sigma$
bands. $\xi_{1}$ is of the order $10^{-3}meV$, since $\Delta$ is
of order $eV$.

So far we have not used the explicit form of $\mathbf{H}_{\sigma}$
and $\mathbf{T}$ in the above discussions. To derive $\xi_{1}$ and
the SOC analytically, we need more details of
$\mathbf{H}_{\sigma}$ and $\mathbf{T}$. $\mathbf{H}_{\sigma}$ can
be written as
\begin{eqnarray}
  \mathbf{H}_{\sigma}=\left(
  \begin{array}{cc}
    \mathbf{E} & \mathbf{\Sigma} \\
    \mathbf{\Sigma}^{\dagger} & \mathbf{E}
  \end{array}
  \right)\otimes\mathbf{I},
\end{eqnarray}
where $\mathbf{I}$ is the identity matrix for the spin degrees of
freedom. The matrix $\mathbf{E}$ represents the onsite energy of
different atomic orbitals, which can be written as
\begin{eqnarray}
  \mathbf{E}=\left(
  \begin{array}{ccc}
    0 & 0 & 0\\
    0 & 0 & 0\\
    0 & 0 & \Delta_{\varepsilon}
  \end{array}
  \right)
\end{eqnarray}
if we arrange the three $sp^{2}$ hybridized orbitals in the
sequence of $\{2p_{y},2p_{x},2s\}$. Here, $\Delta_{\varepsilon}$
is the energy difference $\varepsilon_{2s}-\varepsilon_{2p}$
between the $2s$ and $2p$ orbitals.  $\mathbf{\Sigma}$ describes
the hopping between the two sublattices in the momentum space. To
give its exact form, we first consider the hopping between the two
adjacent atoms in real space, which can also be described by a
$3\times 3$ matrix. Suppose the two adjacent atoms are placed on
the horizontal $x$-axis, i.e., the bond angle is zero, this
hopping matrix can be written as following
\begin{eqnarray}
  \Sigma_{0}=\left(
  \begin{array}{ccc}
    V_{pp\pi} & 0 & 0 \\
    0 & V_{pp\sigma} & V_{sp\sigma} \\
    0 & V_{sp\sigma} & V_{ss\sigma}
  \end{array}
  \right).
\end{eqnarray}
One can obtain the hopping matrix $\Sigma(\theta)$ for arbitrary
bond angle $\theta$ by a rotation $\mathbf{R}(\theta)$ in the
$xy$ plane as
\begin{eqnarray}
  &&\Sigma(\theta)=\mathbf{R}^{\dagger}(\theta)
  \Sigma_{0}\mathbf{R}(\theta)\nonumber\\
  &&\mathbf{R}(\theta)=\left(
  \begin{array}{ccc}
    \cos\theta & -\sin\theta & 0\\
    \sin\theta & \cos\theta & 0\\
    0 & 0 & 1
  \end{array}
  \right).
\end{eqnarray}
The parameters $V_{pp\pi}$, $V_{pp\sigma}$, $V_{sp\sigma}$ and
$V_{ss\sigma}$ correspond the $\sigma$ or $\pi$ bonds formed by
$2s$ and $2p$ orbitals, whose empirical value can be found in
textbooks, for example, Ref.[\onlinecite{saito1}]. Note that we do
not consider the wavefunction overlap matrix in our TBA scheme for
the sake of simplicity. Then the hopping matrix in the momentum
space reads
\begin{eqnarray}
  \mathbf{\Sigma}(\vec{k})=\sum_{\alpha} \Sigma(\theta_{\alpha})
  e^{i\vec{k}\cdot\vec{d}_{\alpha}},
\end{eqnarray}
where, $\vec{d}_{\alpha}$ with $\alpha=1,2,3$ are the bond vectors
connecting the carbon atom and its three nearest neighbors and
$\theta_{\alpha}$ is the angle between $\vec{d}_{\alpha}$ and
$x$-axis.

For $\mathbf{T}$, as we have described above the spin flip on the
same atom only takes place between the $2p_{z}$ and two in-plane
$2p_{x,y}$ orbitals. A straightforward calculation leads to the
onsite spin flip
\begin{eqnarray}
  \mathbf{T}_{o}=\xi_{0}(-\sigma_{x},\sigma_{y},0),
\end{eqnarray}
with $\sigma_{x,y}$ the Pauli matrices. Then $\mathbf{T}$ can be
written as
\begin{eqnarray}
  \mathbf{T}=\left(
  \begin{array}{cc}
    \mathbf{T}_{o} & 0\\
    0 & \mathbf{T}_{o}
  \end{array}
  \right).
\end{eqnarray}
Notice that there are two $\mathbf{T}_{o}$ terms in the above
matrix corresponding to different sublattices.

Since $\mathbf{H}_{\sigma}$ has a large gap near the Dirac points
$K$ and $K^*$, we can expect that
$\mathbf{H}_{\sigma}(\vec{k}+\vec{K})=\mathbf{H}_{\sigma}(\vec{K}) +o(k)$,
which means we can substitute $\mathbf{H}^{-1}_{\sigma}(\vec{K})$ into
Eq.\~(\ref{soeffective}) as a good approximation. Finally we get
the effect Hamiltonian with SOC at the low energy scale,
\begin{eqnarray}
&&H^{[K]}_{eff}\approx\xi_{1}+ \left(
\begin{array}{cc}
\xi_{1}\sigma_{z}&v_{F}(k_{x}+ik_{y})\\
v_{F}(k_{x}-ik_{y})&-\xi_{1}\sigma_{z}
\end{array}
\right)\nonumber\\
&&H^{[K^{*}]}_{eff}\approx\xi_{1}+ \left(
\begin{array}{cc}
-\xi_{1}\sigma_{z}&v_{F}(k_{x}-ik_{y})\\
v_{F}(k_{x}+ik_{y})&\xi_{1}\sigma_{z}
\end{array}
\right). \label{soeff}
\end{eqnarray}
The off-diagonal terms in the above equations come from the
well-known form of $\mathbf{H}_{\pi}$, and $v_{F}$ is just the
Fermi velocity of $\pi$ electrons at the Dirac points. The
effective SOC $\xi_{1}$ in our TBA scheme has an explicit form
\begin{equation}
\xi_{1} \approx |\xi_{0}|^{2}(2\Delta_{\varepsilon})
/(9V^{2}_{sp\sigma}) \label{key-result}
\end{equation}
Eq.(\ref{key-result}) is the key result from our tight-binding
calculation.  Eq.(\ref{soeff}) leads to a spectrum
$E(\vec{k})=\pm\sqrt{(v_{F}k)^{2}+\xi^{2}_{1}}$.  Taking the
values of the corresponding parameters from
Ref.[\onlinecite{saito1}], one can estimate $\xi_{1}$ is of order
$10^{-3}meV$, so is the energy gap $2\xi_{1}$ at the Dirac points.

Eqs. (\ref{soeff}) are similar to those in Ref.
[\onlinecite{kane2005a}], except that the SOC constant $\xi_1$ is
three orders of magnitude smaller than their estimate. We can also
consider the SOC of $\pi$ orbitals between next nearest neighbors
(NNN) which is not forbidden by the symmetry. In this case the
electron moving between NNN will be accelerated by the atoms other
than these two NNN ones which provides the corresponding SOC. This
will involve three center integrals, i.e., two orbital centers and
a potential center which are different with each other. Generally
speaking, such integrals are very small which leads to the SOC of
order at most $10^{-3}meV$ by our estimate and may actually be
smaller.

The argument above is supported by accurate first-principles
calculations based on density-functional theory (DFT). The
relativistic electronic structure of graphene was calculated
self-consistently by the plane-wave method \cite{fang2002} using
the relativistic fully separable pseudopotential in the framework
of noncollinear magnetism \cite{theurich2001}. The
exchange-correlation potential is treated by the local density
approximation (LDA) whose validity for the system considered here
has been demonstrated by many other studies. The experimental
lattice parameter $a=2.456$\AA\ is used in the calculation. The
convergence of calculated results with respect to the number of
{\bf k} points and the cut-off energy has been carefully checked.

\begin{figure}[!htb]
\includegraphics[scale=0.6]{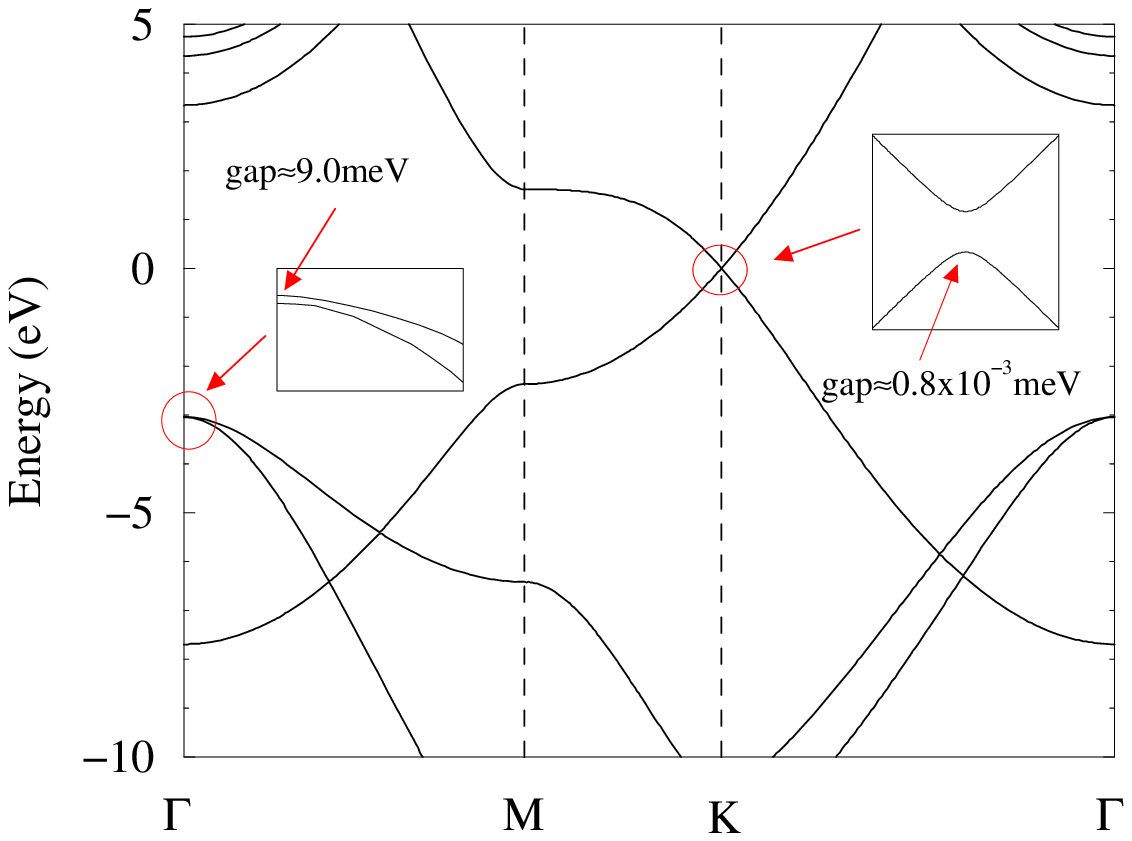}
\caption{(color online). Calculated relativistic band structure of
graphene.}
\end{figure}

Fig. 1 shows the band structure of graphene, we can find that gap
induced by SOC for $\sigma$ orbit is 9.0 meV at $\Gamma$ point.
The figure also indicates that there is a gap induced by SOC for
$\pi$ orbit at K point, and the magnitude of the splitting gap is
$0.8\times 10^{-3}$meV, which is in good agreement with the
estimate obtained from the tight-binding model discussed above.
Since the number discussed here is so small, a few notes are
necessary: 1) the calculations are valid within LDA; 2) the
numeric accuracy of present calculations reachs $10^{-6}$ meV per
atom; 3) the convergency of gap size with respect to the number of
{\bf k} points and cut-off energy is better than $1\times
10^{-4}$meV; 4) the Kramer doublet degeneracy can be reproduced
down to $10^{-5}$meV. Nevertheless, it is clear that the gap
induced by SOC at the K point is of order $10^{-3}$meV.
Considering that graphene may be typically deposited on
substrates, the graphene are generally strained due to small
lattice mismatches, thus the lattice mismatch strain can tune the
splitting gap at K point. We have calculated band structure for
different lattice constants of graphene, and have found that the
splitting gap increases a little with compression while the gap
decreases with tensile strain.

In conclusion, we provided a careful calculation on spin-orbit gap
of graphene, which leads to the same mass term for the
relativistic Dirac fermions in the continuum limit
\cite{kane2005a}, but with a much smaller magnitude of the gap
$10^{-3}{\rm meV}$. The physical reason of the smallness of the
spin-orbit gap can also be understood from the tight-binding model
as coming from the lattice $C_3$ symmetry, which leads to the
vanishing of the leading order contributions. Such a small gap is
consistent with the experimental observation of semi-metallic
behavior of graphene. It shows that the proposed quantum spin Hall
effect in graphene cannot be observed until temperature as low as
$T\ll 10^{-2}K$. In addition, impurity scattering in a disordered
sample may also destabilize the effect.

This work is supported by the Knowledge Innovation Project of the
Chinese Academy of Sciences, the NSFC under the grant numbers
10404035, 10534030 (for Y.G.Y), 90303022, 60576058, 10425418 (for
Z.F.) and 10374058, 90403016 (for X.L.Q.). S. C. Z. is supported
by the NSF through the grant DMR-0342832 and by the US Department
of Energy, Office of Basic Energy Sciences under contract
DE-AC03-76SF00515. We would like to acknowledge helpful
discussions with C. Kane, A. MacDonald, Z. B. Su and B. S. Wang.

\bibliography{Graphene}

\begin{thebibliography}{14}
\expandafter\ifx\csname natexlab\endcsname\relax\def\natexlab#1{#1}\fi
\expandafter\ifx\csname bibnamefont\endcsname\relax
  \def\bibnamefont#1{#1}\fi
\expandafter\ifx\csname bibfnamefont\endcsname\relax
  \def\bibfnamefont#1{#1}\fi
\expandafter\ifx\csname citenamefont\endcsname\relax
  \def\citenamefont#1{#1}\fi
\expandafter\ifx\csname url\endcsname\relax
  \def\url#1{\texttt{#1}}\fi
\expandafter\ifx\csname urlprefix\endcsname\relax\def\urlprefix{URL }\fi
\providecommand{\bibinfo}[2]{#2}
\providecommand{\eprint}[2][]{\url{#2}}

\bibitem[{\citenamefont{Semenoff}(1984)}]{semenoff1984}
\bibinfo{author}{\bibfnamefont{G.~W.} \bibnamefont{Semenoff}},
  \bibinfo{journal}{Phys. Rev. Lett.} \textbf{\bibinfo{volume}{53}},
  \bibinfo{pages}{2449} (\bibinfo{year}{1984}).

\bibitem[{\citenamefont{Novoselov et~al.}(2005)\citenamefont{Novoselov, Geim,
  Morozov, Jiang, Katsnelson, Grigorieva, Dubonos, and Firsov}}]{novoselov2005}
\bibinfo{author}{\bibfnamefont{K.~S.} \bibnamefont{Novoselov}},
  \bibinfo{author}{\bibfnamefont{A.~K.} \bibnamefont{Geim}},
  \bibinfo{author}{\bibfnamefont{S.~V.} \bibnamefont{Morozov}},
  \bibinfo{author}{\bibfnamefont{D.}~\bibnamefont{Jiang}},
  \bibinfo{author}{\bibfnamefont{M.~I.} \bibnamefont{Katsnelson}},
  \bibinfo{author}{\bibfnamefont{I.~V.} \bibnamefont{Grigorieva}},
  \bibinfo{author}{\bibfnamefont{S.~V.} \bibnamefont{Dubonos}},
  \bibnamefont{and} \bibinfo{author}{\bibfnamefont{A.~A.}
  \bibnamefont{Firsov}}, \bibinfo{journal}{Nature}
  \textbf{\bibinfo{volume}{438}}, \bibinfo{pages}{197} (\bibinfo{year}{2005}).

\bibitem[{\citenamefont{Zhang et~al.}(2005)\citenamefont{Zhang, Tan, Stormer,
  and Kim}}]{zhang2005}
\bibinfo{author}{\bibfnamefont{Y.}~\bibnamefont{Zhang}},
  \bibinfo{author}{\bibfnamefont{Y.-W.} \bibnamefont{Tan}},
  \bibinfo{author}{\bibfnamefont{H.~L.} \bibnamefont{Stormer}},
  \bibnamefont{and} \bibinfo{author}{\bibfnamefont{P.}~\bibnamefont{Kim}},
  \bibinfo{journal}{Nature} \textbf{\bibinfo{volume}{438}},
  \bibinfo{pages}{201} (\bibinfo{year}{2005}).

\bibitem[{\citenamefont{Gusynin and Sharapov}(2005)}]{gusynin2005a}
\bibinfo{author}{\bibfnamefont{V.~P.} \bibnamefont{Gusynin}} \bibnamefont{and}
  \bibinfo{author}{\bibfnamefont{S.~G.} \bibnamefont{Sharapov}},
  \bibinfo{journal}{Phys. Rev. Lett.} \textbf{\bibinfo{volume}{95}},
  \bibinfo{pages}{146801} (\bibinfo{year}{2005}).

\bibitem[{\citenamefont{Gusynin and Sharapov}()}]{gusynin2005b}
\bibinfo{author}{\bibfnamefont{V.}~\bibnamefont{Gusynin}} \bibnamefont{and}
  \bibinfo{author}{\bibfnamefont{S.}~\bibnamefont{Sharapov}},
  \bibinfo{howpublished}{cond-mat/0512157}.

\bibitem[{\citenamefont{Peres et~al.}(2006)\citenamefont{Peres, Guinea, and
  Neto}}]{peres2006}
\bibinfo{author}{\bibfnamefont{N.~M.~R.} \bibnamefont{Peres}},
  \bibinfo{author}{\bibfnamefont{F.}~\bibnamefont{Guinea}}, \bibnamefont{and}
  \bibinfo{author}{\bibfnamefont{A.~H.~C.} \bibnamefont{Neto}},
  \bibinfo{journal}{Phys. Rev. B} \textbf{\bibinfo{volume}{73}},
  \bibinfo{pages}{125411} (\bibinfo{year}{2006}).

\bibitem[{\citenamefont{Morozov et~al.}()\citenamefont{Morozov, Novoselov,
  Katsnelson, Schedin, Jiang, and Geim}}]{morozov2006}
\bibinfo{author}{\bibfnamefont{S.}~\bibnamefont{Morozov}},
  \bibinfo{author}{\bibfnamefont{K.}~\bibnamefont{Novoselov}},
  \bibinfo{author}{\bibfnamefont{M.}~\bibnamefont{Katsnelson}},
  \bibinfo{author}{\bibfnamefont{F.}~\bibnamefont{Schedin}},
  \bibinfo{author}{\bibfnamefont{D.}~\bibnamefont{Jiang}}, \bibnamefont{and}
  \bibinfo{author}{\bibfnamefont{A.}~\bibnamefont{Geim}},
  \bibinfo{howpublished}{cond-mat/0603826}.

\bibitem[{\citenamefont{Kane and Mele}(2005)}]{kane2005a}
\bibinfo{author}{\bibfnamefont{C.~L.} \bibnamefont{Kane}} \bibnamefont{and}
  \bibinfo{author}{\bibfnamefont{E.~J.} \bibnamefont{Mele}},
  \bibinfo{journal}{Phys. Rev. Lett.} \textbf{\bibinfo{volume}{95}},
  \bibinfo{pages}{226801} (\bibinfo{year}{2005}).

\bibitem[{\citenamefont{Murakami et~al.}(2004)\citenamefont{Murakami, Nagaosa,
  and Zhang}}]{murakami2004B}
\bibinfo{author}{\bibfnamefont{S.}~\bibnamefont{Murakami}},
  \bibinfo{author}{\bibfnamefont{N.}~\bibnamefont{Nagaosa}}, \bibnamefont{and}
  \bibinfo{author}{\bibfnamefont{S.-C.} \bibnamefont{Zhang}},
  \bibinfo{journal}{Phys. Rev. Lett.} \textbf{\bibinfo{volume}{93}},
  \bibinfo{pages}{156804} (\bibinfo{year}{2004}).

\bibitem[{\citenamefont{Haldane}(1988)}]{haldane1988}
\bibinfo{author}{\bibfnamefont{F.~D.~M.} \bibnamefont{Haldane}},
  \bibinfo{journal}{Phys. Rev. Lett.} \textbf{\bibinfo{volume}{61}},
  \bibinfo{pages}{2015} (\bibinfo{year}{1988}).

\bibitem[{\citenamefont{Saito et~al.}(1998)\citenamefont{Saito, Dresselhaus,
  and Dresselhaus}}]{saito1}
\bibinfo{author}{\bibfnamefont{R.}~\bibnamefont{Saito}},
  \bibinfo{author}{\bibfnamefont{G.}~\bibnamefont{Dresselhaus}},
  \bibnamefont{and} \bibinfo{author}{\bibfnamefont{M.~S.}
  \bibnamefont{Dresselhaus}}, \emph{\bibinfo{title}{Physical Properties of
  Carbon Nanotubes}} (\bibinfo{publisher}{Imperial College Press},
  \bibinfo{address}{London}, \bibinfo{year}{1998}).

\bibitem[{\citenamefont{Ando}(2000)}]{ando2000}
\bibinfo{author}{\bibfnamefont{T.}~\bibnamefont{Ando}}, \bibinfo{journal}{J.
  Phys. Soc. Japan} \textbf{\bibinfo{volume}{69}}, \bibinfo{pages}{1757}
  (\bibinfo{year}{2000}).

\bibitem[{\citenamefont{Fang and Terakura}(2002)}]{fang2002}
\bibinfo{author}{\bibfnamefont{Z.}~\bibnamefont{Fang}} \bibnamefont{and}
  \bibinfo{author}{\bibfnamefont{K.}~\bibnamefont{Terakura}},
  \bibinfo{journal}{J. Phys.: Condens. Matter} \textbf{\bibinfo{volume}{14}},
  \bibinfo{pages}{3001} (\bibinfo{year}{2002}).

\bibitem[{\citenamefont{Theurich and Hill}(2001)}]{theurich2001}
\bibinfo{author}{\bibfnamefont{G.}~\bibnamefont{Theurich}} \bibnamefont{and}
  \bibinfo{author}{\bibfnamefont{N.}~\bibnamefont{Hill}},
  \bibinfo{journal}{Phys. Rev. B} \textbf{\bibinfo{volume}{64}},
  \bibinfo{pages}{073106} (\bibinfo{year}{2001}).

\end{thebibliography}

\end{document}